# Preparation sequence controls oxygen partitioning between boron and lithium on tungsten

Predrag S Krstic[1]*, Meral Sharkass[2], Swarit Dwivedi[3], Dalia Sayed Ahmed[2], Yun Kyung Shin[4], and Adri CT van Duin[4]

*1 TheoretiK Consulting, Ossining, New York 10562, USA*
*2 Department of Nuclear Engineering, The Pennsylvania State University, University Park, Pennsylvania 16801, USA*
*3 Department of Materials Science and Engineering, Monash University, Clayton, Victoria, 3800, Australia*
*4 Department of Mechanical Engineering, The Pennsylvania State University, University Park, Pennsylvania 16802, United States*

## Abstract

Classical reactive molecular dynamics of boron- and lithium-conditioned tungsten surfaces show that the order in which boron and oxygen are introduced - not the boron inventory - controls surface chemistry. On a thin engineered tungsten oxide, boron captures 81% of oxygen and consumes the oxide; on an engineered ~2 nm mixed $WO_3/WO_2$ film, boron captures only 20% and leaves 52% of reaction-zone tungsten in $WO_3$-like coordination; and in a co-deposited W-B substrate later exposed to oxygen, boron captures only 28% despite a tenfold larger inventory. Adding lithium leaves the oxide-first cases unchanged but transforms the co-deposited case, where lithium becomes the dominant oxygen captor (58%) and a lithium-borate-glass-like network form. We outline the expected, divergent responses of the boron-only and boron-plus-lithium materials to deuterium irradiation.

## 1. Introduction

Boronization and lithiation are routinely used to condition the plasma-facing components (PFCs) of magnetic-confinement fusion devices, suppressing intrinsic impurities, gettering oxygen and controlling fuel recycling [1-6]. Originally developed for carbon and beryllium walls, these low-Z conditioning techniques have had to be reassessed as the field has converged on tungsten as the reference PFC material. Real-time boron and boron-nitride powder injection now supplements conventional glow-discharge boronization on the full-tungsten walls of ASDEX Upgrade and EAST, where it reduces tungsten and oxygen influx and recovers high-performance scenarios between boronizations [1-4]. Surface analyses of these machines show that injected boron migrates and redeposits as a mixed tungsten-boron layer in erosion-dominated regions and as closed boron films elsewhere [3]. With the elimination of beryllium in the re-baselined ITER, boron coatings have become central to wall conditioning: a diborane glow-discharge boronization system is foreseen primarily to getter oxygen and limit tungsten sputtering during the limiter start-up phase, and the lifetime, migration and fuel-retention consequences of these coatings are now active questions for ITER and DEMO [7-10]. In parallel, lithium conditioning remains attractive for its strong control of hydrogenic recycling, and boron and lithium have been applied together on several devices. On carbon-wall machines the two were combined early: solid lithium and boron pellets were both injected into TFTR, each improving wall conditioning and supershot performance [11], and the spherical tokamak NSTX (later NSTX-U) routinely combined trimethyl-boron boronization with lithium evaporation on its graphite plasma-facing components [12]. On metal-wall machines the same pairing is now being revisited: the full-metal-wall tokamaks EAST and ASDEX Upgrade both employ boronization together with lithium conditioning, EAST having coated its molybdenum first wall and carbon divertor with lithium and, more recently, boronized an ITER-like tungsten divertor [1-5]. Whether boron and lithium play the same chemical roles on tungsten as on carbon is, however, far from obvious. The conditioning effect of boron and lithium is, at bottom, a surface-chemistry problem: combined experimental and atomistic studies of lithiated, boronized and oxidized carbon surfaces have shown that oxygen, gettered from residual water vapor and structural steels, rather than the alkali or low-Z getter itself, frequently controls the binding of hydrogenic fuel [13-16]. Whether the same oxygen-centric picture governs conditioned tungsten surfaces, and how it depends on how the coating is built, remains largely unexplored. The atomic-scale outcome may in fact depend on preparation route as much as on composition. We compare three routes (Figure 1) - boron deposited on a thin engineered oxide (OX-THIN), on a grown ~2 nm mixed WO3/WO2 film (OX-THICK), and boron engineered into W before oxygen irradiation (CO-DEP) - each analyzed both before and after lithium deposition. The oxidation of tungsten and the behavior of oxygen at tungsten surfaces have been examined previously by molecular dynamics, including early simulations of oxygen-driven tungsten erosion under deuterium bombardment [17], reactive-forcefield studies of low-energy oxygen-plasma oxidation [18], and machine-learned-potential studies of particle impact on tungsten-oxide surfaces [19]; the incorporation and migration of boron within tungsten have likewise been addressed atomistically [20]. What has not been examined is how the sequence in which boron and oxygen are brought together on a tungsten surface determines the resulting chemistry. Here we show, using classical reactive molecular dynamics, that this preparation route - not the boron inventory - governs whether boron consumes oxide, whether tungsten remains oxidized, and whether subsequently deposited lithium is inert or instead forms a lithium-borate-glass-like network. We then outline the divergent responses these distinct materials should display under deuterium irradiation.

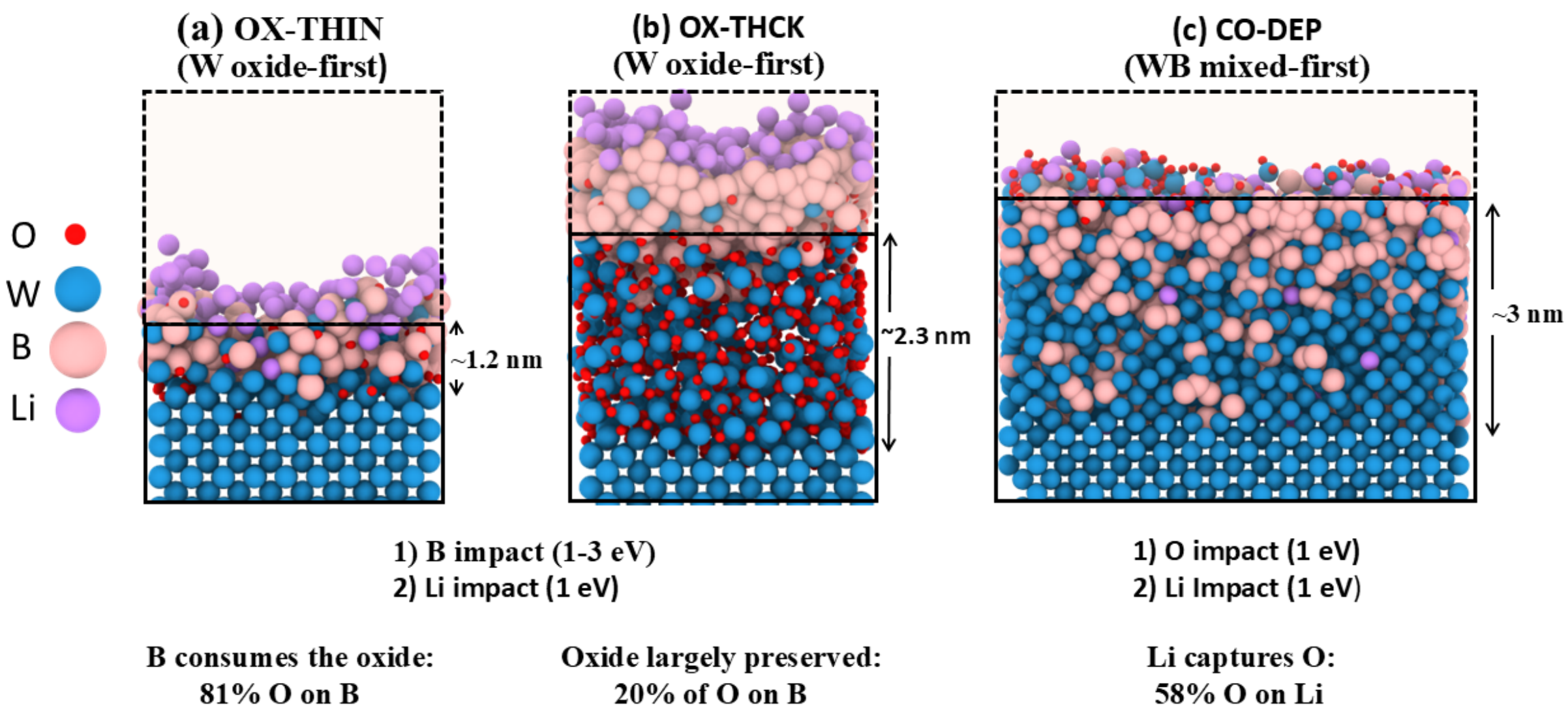

**Figure 1.** Final near-surface structure of the three preparation routes (surface-aligned schematic, consistent with the density profiles of Figure 3). Each panel shows the W(BCC) substrate, the oxide or boron layer with its approximate thickness, and the embedded O, B and Li produced by the preparation sequence noted beneath the panel. (a) OX-THIN (W oxide-first): a thin graded W-O oxide (~1.2 nm) is boronized and then lithiated; boron consumes the oxide, capturing 81% of the oxygen, and lithium remains a near-surface spectator. (b) OX-THICK (W oxide-first): a grown ~2.3 nm $WO_3/WO_2$ film is boronized and lithiated; the oxide is largely preserved, boron captures only 20% of the oxygen, and lithium is excluded. (c) CO-DEP (mixed WB-first): a graded W-B layer (~3 nm) is irradiated by oxygen and then lithium; boron is broadly distributed deep in the metal while oxygen and lithium remain near the surface, where lithium captures 58% of the oxygen and forms a lithium-borate network. Dimensions of atoms are chosen for better visibility. All irradiations are monoenergetic and at normal incidence (along -z); the full construction protocol lis given in the Supplemental Material.

## 2. Methods

All simulations were performed using classical reactive molecular dynamics with a newly developed W/O/Li/B/D ReaxFF parameter set. ReaxFF provides a flexible and transferable interatomic potential capable of describing bond formation and dissociation in large, chemically complex systems over timescales relevant to surface conditioning [21,22]. This force field was developed by reparameterizing the previously reported Li/B/O/D force filed [15, 23] and integrating it with the W/O/D force field [18]. The combined description was subsequently extended to include previously unparameterized cross-interactions among the Li-, B-, and W-based systems; the resulting formulation will be presented in a separate publication.

The three substrates were built to mimic the distinct experimental histories of interest. The two oxide-first substrates (OX-THIN, OX-THICK) begin from a crystalline tungsten slab carrying the thin and thick oxides introduced above; boron and then lithium are deposited onto the oxidized surface. CO-DEP is built in the opposite order: boron is engineered into tungsten with a graded profile (about 70% B at the surface, decaying to zero near 3 nm depth), and only afterwards is oxygen introduced by 1 eV irradiation, followed by lithium. The three slabs contain 5,328-16,758 atoms (Table 1); each was relaxed and thermalized to room temperature before analysis, and quantities are reported from the final equilibrated configuration. All depositions use monoenergetic atoms at normal incidence; boron is deposited at 3 eV, oxygen at 1 eV, and lithium with a 1 eV then 3 eV two-stage schedule for the oxide-first routes and at 1 eV for CO-DEP. The full per-route construction and relaxation protocol is given in the Supplemental Material.

Bonding was characterized from the ten partial radial distribution functions and from cutoff-based coordination numbers for every species pair, supplemented by a connected-component cluster analysis that classifies each boron and lithium atom by its local environment (e.g. interfacial, metallic-cluster, or oxygen-coordinated) and each oxygen atom by the set of metals it bridges. Because the slabs are dominated by bulk metal - only 6-22% of the tungsten lies in the near-surface region where boron, oxygen and lithium reside - averaging over all tungsten badly dilutes the surface signal. We therefore define a reaction zone extending from the deepest non-tungsten atom plus a 2 Å margin to the surface, and report tungsten-oxidation statistics over this zone; doing so raises the apparent W-O coordination by 4-16 times relative to whole-slab averages and is essential for a meaningful comparison across routes. Depth profiles are surface aligned so that z = 0 marks the original pre-treatment surface, allowing the three histories to be overlaid directly.

## 3. Results and discussion

**Boron-only chemistry.** The three routes give three regimes (Table 1). In OX-THIN, boron reaches and consumes the thin oxide - 81% of oxygen moves to boron and the reaction-zone tungsten is left near-metallic. In OX-THICK, the structured oxide resists boronization: only 20% of oxygen moves to boron and 52% of reaction-zone tungsten keeps $WO_3$-like coordination, the W-O bond length matching crystalline oxide. In CO-DEP, despite fivefold more boron per oxygen than OX-THIN, only 28% of oxygen bonds to boron, because most boron is buried in the tungsten matrix and chemically inaccessible.

**Table 1.** Key observables across the three routes (reaction-zone W for the tungsten-state block).

| Quantity | OX-THIN | OX-THICK | CO-DEP |
|---|---|---|---|
| *Composition* | | | |
| N(O) / N(B) / N(Li) | 172/219/98 | 991/466/128 | 305/2007/201 |
| B/O ratio | 1.27 | 0.47 | 6.58 |
| Preparation route | oxide-first | oxide-first | mixed-first |
| *Oxygen capture* | | | |
| O bonded to B [%] | 80.8 | 20.1 | 28.2 |
| O bonded to Li [%] | 2.3 | 0.1 | 58.0 |
| *Tungsten state (reaction zone)* | | | |
| $WO_3$-like W (≥5 O) [%] | 0.3 | 51.8 | 1.0 |
| Avg O per W | 0.81 | 3.59 | 0.14 |
| *Lithium-borate markers* | | | |
| O in W-B-Li sites | 3 | 1 | 18 |
| Li with O coordination | 5 | 1 | 127 |

**Lithium reverses the hierarchy.** Lithium deposition barely perturbs the oxide-first systems - boron has already taken the accessible oxygen, so 95-99% of lithium is isolated and ≤2.3% of oxygen bonds to lithium. In CO-DEP the opposite occurs: lithium captures 58% of oxygen versus 28% for boron (Figure 2), the dominant motif becomes W-Li bridging, and 18 oxygen atoms occupy W-B-Li network sites - the signature of a lithium-borate glass. Density profiles confirm the structural distinction (Figure 3): in CO-DEP boron is distributed deep into the tungsten matrix while lithium and oxygen remain concentrated near the surface, whereas in the oxide-first systems oxygen extends deepest with boron and lithium above it.

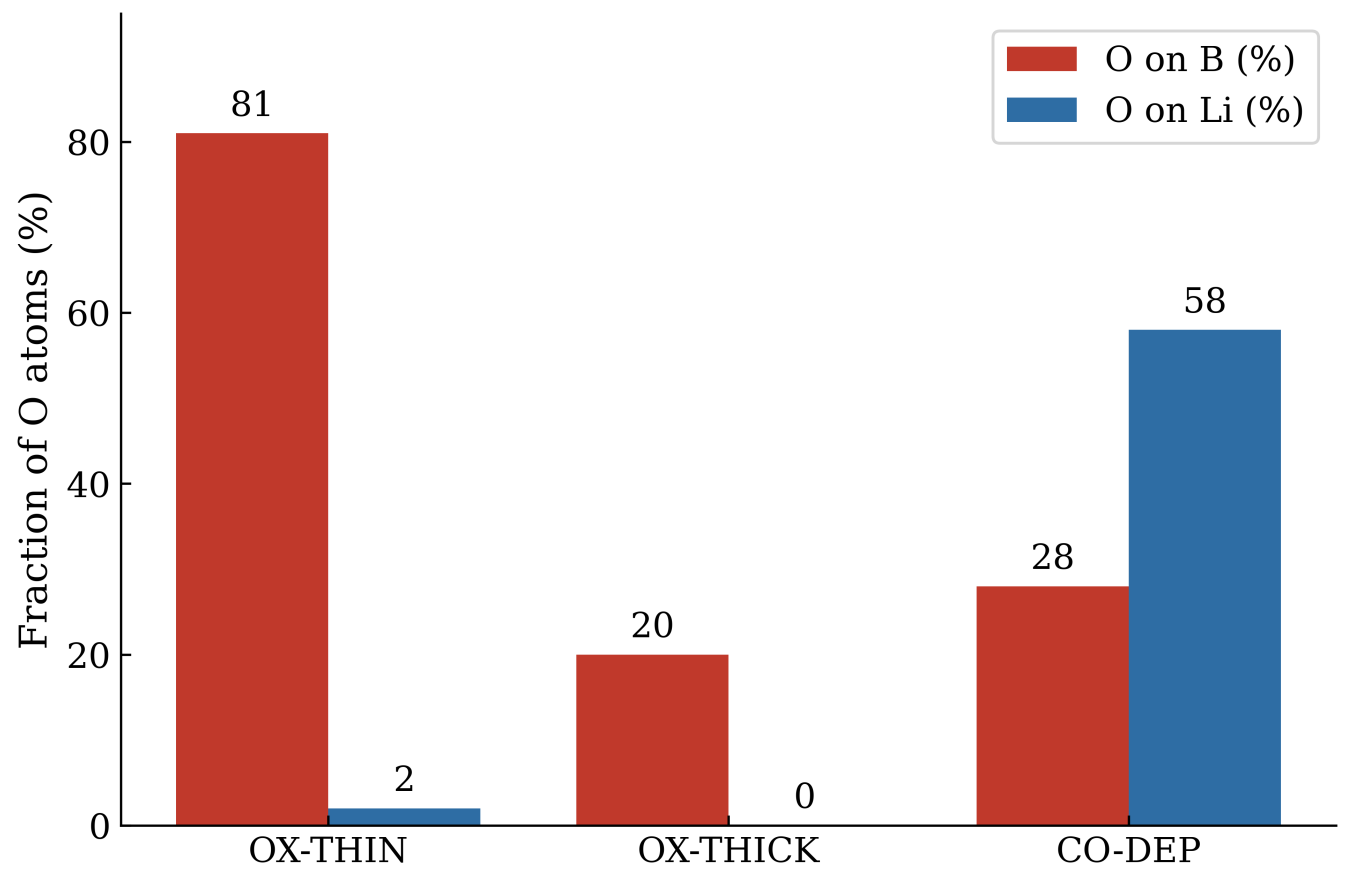

**Figure 2.** Oxygen captor, boron versus lithium: fraction of oxygen atoms bonded to boron (red) and to lithium (blue) across the three routes. Boron dominates in the oxide-first systems; lithium dominates in CO-DEP.

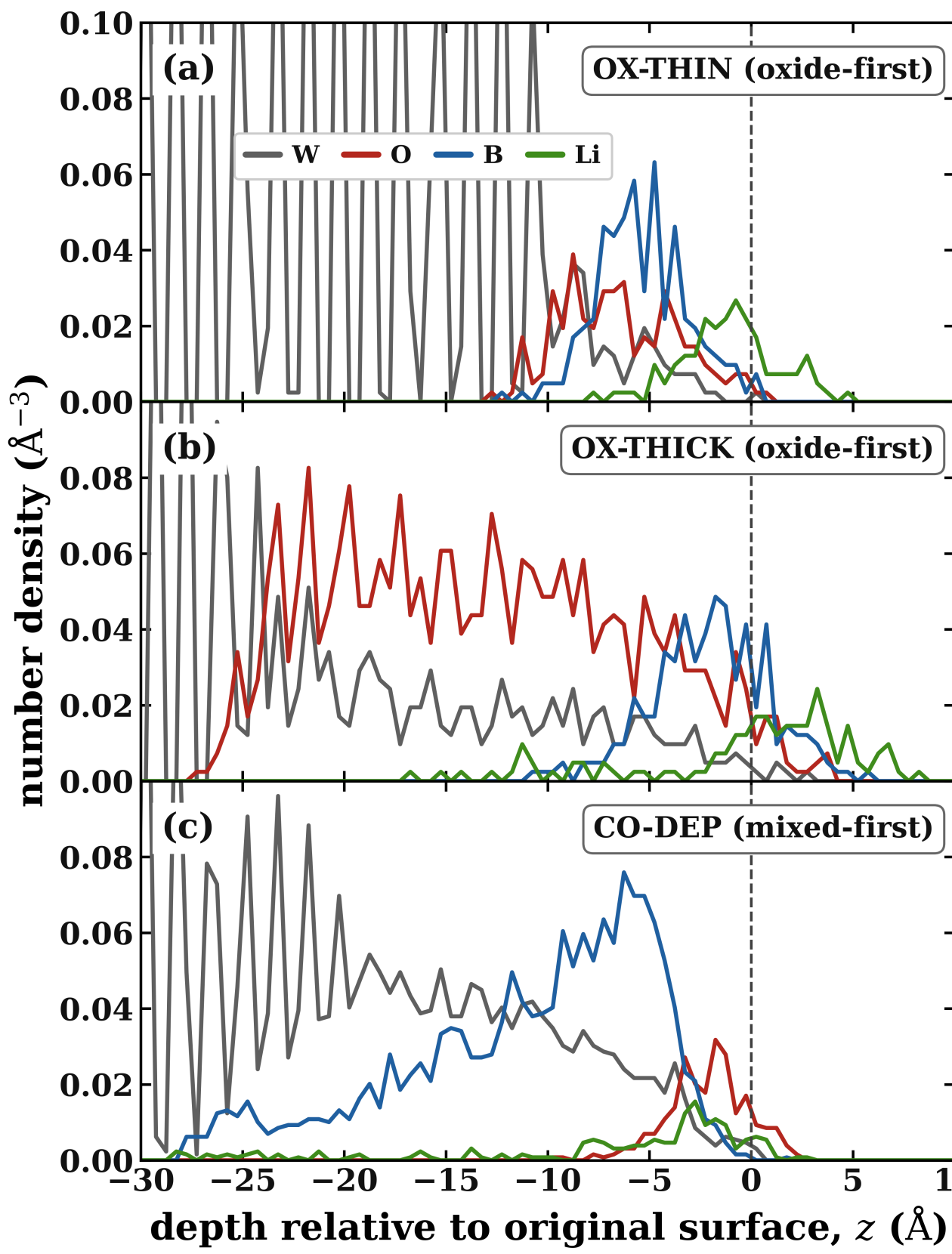

**Figure 3.** Atomic density profiles of the W-O-B-Li system, surface-aligned with z = 0 at the original surface (dashed line) and depth increasing to the left. OX-THIN: thin O-B layer on metallic W. OX-THICK: thick preserved oxide. CO-DEP: B buried deep in metallic W, with O and Li shallow.

**Process route over elemental ratio.** The central result is that boron abundance does not predict reactivity: CO-DEP has roughly fivefold more boron per oxygen than OX-THIN (B/O = 6.58 versus 1.27) yet captures only about a third as much oxygen on boron (28% versus 81%), because preparation sequence fixes where boron sits relative to oxygen. Practically, the conventional "boronize then lithiate" order on a pre-oxidized surface suppresses lithium uptake, whereas co-deposition enables a lithium-rich glassy borate phase. Bonded O-Li (~2.0 Å) and B-Li (~2.4 Å) RDF peaks appear only in CO-DEP (Figure 4), confirming lithium-borate formation. This resolves the apparent paradox that the most boron-rich surface is among the least efficient at boronizing its oxygen: only boron presented to oxygen at the right stage participates, while boron buried beforehand is sequestered in B-B bonds. Lithium succeeds in CO-DEP not by thermodynamic preference over boron but because it arrives with the oxygen and tolerates the irregular, low-coordination sites of a metallic W-B substrate, where ordered borate units cannot assemble.

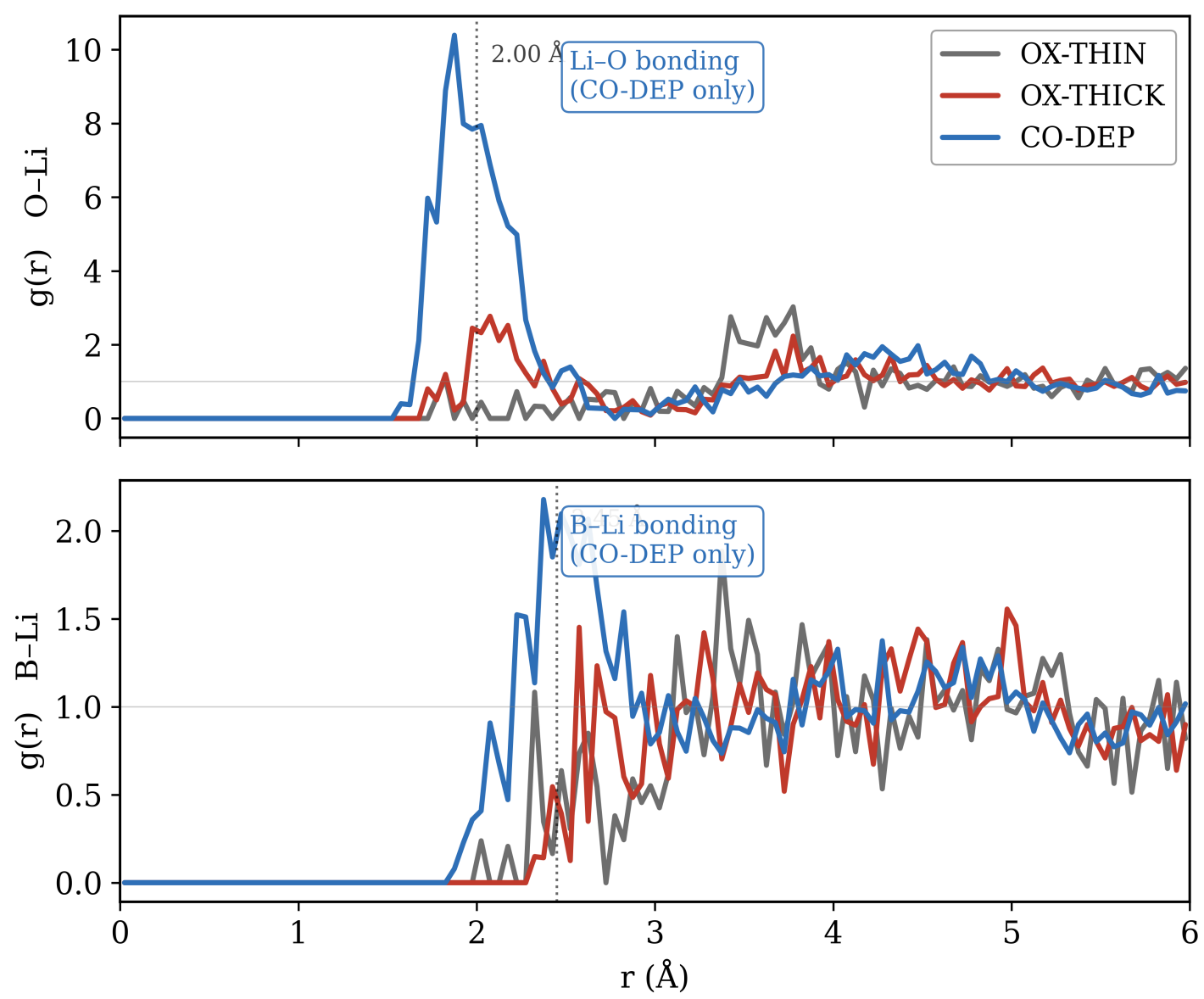

**Figure 4.** Tail-normalized O-Li and B-Li partial radial distribution functions across the three routes. A bonded O-Li first peak near 2.0 Å and a B-Li first peak near 2.4 Å (dotted reference lines) develop only in CO-DEP (blue); in OX-THIN (grey) and OX-THICK (red) both pairs remain at or below g(r) = 1, indicating no Li-O or Li-B bonding. The grey horizontal line marks g(r) = 1. These peaks are the bond-length-resolved signature of the lithium-borate network (the OX-THIN and OX-THICK O-Li signals are statistically negligible; see Table 1).

## 4. Implications for deuterium irradiation

The boron-only and boron-plus-lithium materials should differ under D irradiation. The retention of deuterium in boronized, lithiated and oxidized surfaces, and the controlling role of oxygen in that chemistry, have been established in earlier combined experimental and atomistic studies of low-Z conditioned carbon, where deuterium was found to bind preferentially at oxygen and oxygen-carbon sites rather than at the getter atoms themselves [13-16]. Carrying that oxygen-centric picture over to tungsten, the three routes here should respond very differently. For the boron-only material: OX-THIN retains D as D-B and D-O in a thin reacted layer; OX-THICK retains D in the preserved oxide, with the intact $WO_3$/$WO_2$ film acting as a barrier that limits D transport to the metallic tungsten beneath; and CO-DEP allows deeper D penetration along the buried boron network, with a larger metallic-tungsten sputtering contribution because the surface is not oxide-protected [20].

Adding lithium changes both retention and erosion. In OX-THIN and OX-THICK the lithium forms a weakly bound, near-metallic surface cap that is expected to be preferentially sputtered at the lowest D energies, exposing the boron or oxide layers beneath while leaving the underlying D chemistry essentially as in the boron-only case. In CO-DEP the lithium-borate network is expected to dominate the response: lithium-borate phases are known lithium-ion conductors and may retain deuterium as Li-D or within D-O-B network sites, plausibly giving the highest D retention of the three routes, while lithium's exposed near-surface position should give the highest lithium sputtering yield. The contrast between the boron-only and lithium-bearing versions of CO-DEP - metallic-network B-D retention versus glassy lithium-borate D retention - is the sharpest prediction to test. Quantifying these trends through a systematic 1-100 eV deuterium campaign, with per-species sputtering yields and depth-resolved D speciation, is the natural next step and will be the subject of a forthcoming publication.

## 5. Conclusions

Using reactive molecular dynamics, we have shown that the preparation route - the order in which boron and oxygen meet on a tungsten surface - controls the resulting surface chemistry more strongly than the boron inventory does. The boron-only material shows three regimes: clean boronization that consumes a thin native oxide (OX-THIN), boronization arrested by a structured oxide that survives largely intact (OX-THICK), and deep boron incorporation that leaves most boron sequestered and the oxide barely reduced (CO-DEP). Lithium then acts as a sensitive probe of how much oxygen remains available: it is essentially inert in the oxide-first routes, where boron has already taken the accessible oxygen, but in the co-deposited route it outcompetes boron for oxygen and forms a lithium-borate-glass-like network on metallic tungsten.

These findings carry a concrete message for wall conditioning of tungsten devices: simply increasing the boron inventory or applying boron and lithium in the conventional sequence on an already-oxidized surface, need not produce the intended passivation and may suppress the lithium chemistry altogether; the timing of boron, oxygen and lithium delivery is itself a design variable. Because the three routes yield chemically and structurally distinct surfaces - a thin reacted layer, an intact oxide barrier, and a lithium-borate glass on metallic W-B - they should differ markedly in deuterium retention and sputtering, with direct consequences for fuel inventory and impurity influx in ITER- and DEMO-relevant scenarios. Establishing those differences quantitatively under deuterium irradiation is the immediate next objective.

### *Acknowledgments*

This work is supported by the U.S. Department of Energy, Office of Science/Fusion Energy Sciences under Award Number DE-SC0022013 to Pennsylvania State University. PK, MS and DSA acknowledge use of SDSC Expanse HPC facility, enabled by ACCESS grant TG-DMR110037 through ORNL. PK is grateful to Dr. Camilo Jaramilo Correa and Hanna Schamis of PPPL for inspiring discussions.

### *Data Availability Statement*

The molecular dynamics simulation datasets, interatomic potential parameters, surface trajectory files, and input scripts supporting the findings of this study are publicly archived and accessible in the Zenodo repository at https://doi.org/10.5281/zenodo.21494702.

# Supplemental Material

This Supplemental Material gives the full construction protocol for the three W–O–B–Li slabs, sufficient to reproduce the substrates analyzed in the main text. All slabs are built on a crystalline tungsten (BCC) substrate. Every deposition or irradiation step uses monoenergetic atoms launched along the surface normal (−z, perpendicular to the top face) from 10 Å above the surface — beyond the ReaxFF interaction cutoff (~5–6 Å), so the incoming trajectory is ballistic until the atom enters the interaction region. No energy or angular distributions are sampled: in every case the impact energy is fixed and the impact direction is normal incidence. Each construction step is followed by the same relaxation sequence: removal of overlapping atoms, energy minimization, NVT thermalization, NPT lateral relaxation, an NVT production run, and an NVE stability check.

### *S1. OX-THIN — graded thin oxide, then boronization, then lithiation*

The thin oxide is engineered as a compositionally graded W–O layer with a depth-dependent stoichiometry: the top 0–3 Å bin is set to a $WO_3$ target composition, with the oxygen content decreasing in successive bins (3–5 Å, 5–9 Å, 9–13 Å) and zero oxygen beyond $z \approx 13$ Å. After the standard relaxation sequence, the surviving stable oxide contains 172 O atoms over a thickness of ~14 Å, with an average O:W ratio of approximately 2:1 (sub-stoichiometric, a mixed WO/$WO_2$ character). The oxide is then boronized (Sec. S3) and lithiated (Sec. S4).

### *S2. OX-THICK — grown thick oxide, then boronization, then lithiation*

The thick oxide is a mixed $WO_3$/$WO_2$ film of ~2.3 nm, grown by energetic (30 eV) oxygen bombardment as described in Ref. [18]. After relaxation it carries 991 O atoms. It is then boronized (Sec. S3) and lithiated (Sec. S4) by the same protocols as OX-THIN.

### *S3. Boronization of the oxide-first routes (WO → WOB)*

Boron is deposited at 3 eV via a hybrid protocol — a gentle deposition phase followed by a 3 eV impact phase. The 3 eV energy is a deliberate choice: at 1 eV boron scatters off the surface potential without reaching a chemisorption site (only weak physisorption results), whereas above ~10 eV boron penetrates and causes displacement damage, turning boronization into implantation. At 3 eV boron clears the surface potential well and forms B–O and B–W bonds within the top 2–3 Å without damage. Boron atoms are launched from 10 Å above the surface (ballistic approach, as above). OX-THIN and OX-THICK use the same 3 eV energy; only the boron dose differs, giving B/O ≈ 1.27 for OX-THIN and B/O ≈ 0.47 for OX-THICK (Table 1).

### *S4. Lithiation of the oxide-first routes (WOB → WOBLi)*

Lithium is deposited with a two-stage energy schedule: the first half of the lithium batches are delivered at 1 eV (gentle impacts that allow a Li–O reaction layer to form first), and the second half at 3 eV (more energetic impacts). This two-stage 1 eV/3 eV schedule is used only for the oxide-first routes (OX-THIN and OX-THICK), matching the 3 eV energy used to boronize them; CO-DEP instead uses single-energy 1 eV deposition for both oxygen and lithium (Sec. S5).

### *S5. CO-DEP — graded W–B substrate, then oxygen, then lithium*

CO-DEP is built in the reverse order. A compositionally graded W–B layer is engineered with a depth-dependent boron fraction: 70% B in the top 0–10 Å bin, decreasing to 40% B (10–20 Å) and 20% B (20–30 Å), with zero boron beyond z = 30 Å. After the standard relaxation sequence this yields a stable W–B slab. The W–B substrate is then irradiated with 1 eV oxygen atoms and, after a further relaxation sequence, with 1 eV lithium atoms.

### *S6. Common settings*

Interatomic interactions are described by the W/O/Li/B/D ReaxFF parameter set used in the main text (the fixed-energy, normal-incidence impact conditions common to all three routes are stated above). Final atom inventories and cell dimensions for the three slabs are listed in Table 1.